\documentclass[fleqn,12pt,twoside]{article}
\usepackage[headings]{espcrc1}

\usepackage{psfrag}

\readRCS
$Id: espcrc1.tex,v 1.2 2004/02/24 11:22:11 spepping Exp $
\usepackage{graphicx}

\newcommand{\AmS}{{\protect\the\textfont2
  A\kern-.1667em\lower.5ex\hbox{M}\kern-.125emS}}

\title{Fractals Meet Fractals: Self-Avoiding Random Walks on
Percolation Clusters}

\author{Viktoria Blavatska\address{Institute for Condensed Matter Physics,
National Academy of Sciences of Ukraine,\\
1 Svientsitskii Str., 79011 Lviv, Ukraine}$^{{\rm ,b}}$
        and 
        Wolfhard Janke\address[2]{Institut f\"ur Theoretische Physik and Centre for
  Theoretical Sciences (NTZ), Universit\"at Leipzig, Postfach 100\,920,
  04009 Leipzig, Germany}}
       
\runtitle{Fractals Meet Fractals: Self-Avoiding Random Walks on
Percolation Clusters}
\runauthor{Viktoria Blavatska and Wolfhard Janke}

\begin{document}

\maketitle

\begin{abstract}
The scaling behavior of linear polymers in disordered media, modelled by 
self-avoiding walks (SAWs) on the backbone of percolation clusters in two, 
three and four dimensions is studied by numerical simulations. We apply 
the pruned-enriched Rosenbluth chain-growth method (PERM). Our numerical 
results yield estimates of critical exponents, governing the 
scaling laws of disorder averages of the configurational properties 
of SAWs, and clearly indicate a multifractal spectrum which
emerges when two fractals meet each other.
\end{abstract}


\section{INTRODUCTION}

Self-avoiding walks (SAWs) on regular lattices provide a successful
description of the universal configurational  properties of polymer chains
in good solvent \cite{deGennes79,desCloizeaux90}. In particular, the average 
square end-to-end distance $\langle R^2\rangle$ of SAWs with $N$ steps obeys
the scaling law
\begin{equation}\label{scaling}
 \langle R^2 \rangle
\sim N^{2\nu_{{\rm SAW}}},\mbox{\hspace{3em}}
\end{equation} 
where the universal exponent $\nu_{{\rm SAW}} > 1/2$ only depends on the
space dimension $d$. For regular lattices, its value is well established
($\nu_{{\rm SAW}} = 3/4, 0.5882(11), 1/2$ for $d=2,3,\ge 4$). 
New challenges have been raised recently in studies of biopolymers in natural
cellular environments, which are usually very crowded by many other biochemical
species occupying a large fraction of the total volume \cite{crowding}. In a minimalistic 
description, we assume here this ``volume exclusion'' to be random and frozen,
i.e., quenched, and model the available space for the SAWs by site percolation
 clusters on hypercubic lattices at the percolation threshold 
$p_c = 0.592\,746, 0.311\,60, 0.196\,88$ in $d=2,3,4$. Note that a percolation
cluster itself is a fractal object with fractal dimension $d_{p_c}^{B}$ dependent on the 
space dimension $d$. The scaling law (\ref{scaling}) still
holds, but with an exponent $\nu_{p_c}\neq\nu_{{\rm SAW}}$ 
\cite{Woo91,Meir89,Kim83,Barat91,Grassberger93,Lee96,Rintoul94,Ordemann00,Blavatska04,Janssen07,1,2,3}.

When studying physical processes on complicated fractal objects, one often encounters 
the situation of coexistence of a family of singularities, 
each associated with a set of different fractal dimensions \cite{Stanley88}. 
In these problems, an infinite set of critical exponents 
is needed to characterize the different moments of the distribution of observables, which scale independently.
These peculiarities are usually referred to as multifractality \cite{Mandelbrot74}.  
Multifractal properties arise in many different contexts, for example in studies of turbulence in chaotic dynamical systems and 
strange attractors \cite{Mandelbrot74,Hentschel83}, human heartbeat dynamics \cite{Ivanov99}, Anderson localization transition \cite{Schreiber91},
etc. 

Although the behavior of SAWs on percolative lattices served as a subject of numerous numerical  and analytical studies since the early 80th, 
not enough attention has been paid to clarifying the multifractality of the problem. 
It was only recently proven in field-theoretical studies \cite{Blavatska04,Janssen07} 
that the exponent $\nu_{p_c}$ alone is not sufficient to completely describe the peculiarities of SAWs on percolation clusters. 
Instead, a whole spectrum $\nu^{(q)}$ of multifractal exponents emerges \cite{Janssen07}:
 \begin{eqnarray}
\nu^{(q)}{=}\frac{1}{2}{+}\left(\!\frac{5}{2}{-}\frac{3}{2^q}\!\right)\frac{\varepsilon}{42}{+}\left(\!\frac{589}{21}{-}\frac{397}{14\cdot2^q}{+}\frac{9}{4^q} \!\right)\left( \frac{\varepsilon}{42}\right)^2, \label{spectr}
\end{eqnarray} 
with $\varepsilon=6-d$. 
Note that putting $q=0$ in (\ref{spectr}), we restore an estimate for the dimension $d_{p_c}^B$ of the underlying backbone of percolation clusters  via $\nu^{(0)}=1/d_{p_c}^B$, and
$\nu^{(1)}$ gives us the exponent $\nu_{p_c}$, governing the scaling law for the averaged end-to-end distance of SAWs on the backbone of percolation clusters.
In the present paper,
we report a careful computer simulation study of SAWs on percolation clusters.

\section{METHODS}

We consider site percolation on regular lattices of edge lengths up to $L_{{\rm max}}{=}400,200,50$ in dimensions $d{=}2,3,4$, respectively. Each site of 
the lattice is occupied randomly with probability $p_c$ and empty otherwise.
 To  extract 
the backbone of a percolation cluster,  we apply the algorithm  proposed in Ref.~\cite{Porto97}.  

We construct a SAW on percolation clusters, applying the pruned-enriched chain-growth algorithm 
\cite{Rosenbluth55,Grassberger97,Bachmann03}.
 We let a trajectory of SAW grow step by step, until it reaches some prescribed distance (say $R$) from the starting point. Then, the algorithm is stopped,  
and a new SAW grows from the same starting point. In such a way, we are interested in constructing different possible trajectories with fixed end-to-end distance. For each lattice size $L$, we change $R$ up to $\approx L/3$ to avoid finite-size effects, since close to 
lattice borders the structure of the backbone of percolation clusters is distorted and thus can falsify the SAW statistics.   

Let us denote by $K(R)$ the total number of constructed SAW trajectories between $0$ and $R$ (we perform $\sim10^6$ SAWs for each value of $R$). 
Then, for each site $i$ of the backbone  we sum up the portion of trajectories, passing through this site. In such a way, we prescribe a weight $w(i)=K(i)/K(R)$ 
to each site $i\in R$ of the underlying fractal cluster. 
 
 \begin{figure}[t!]
 \begin{center}
\includegraphics[width=5.9cm]{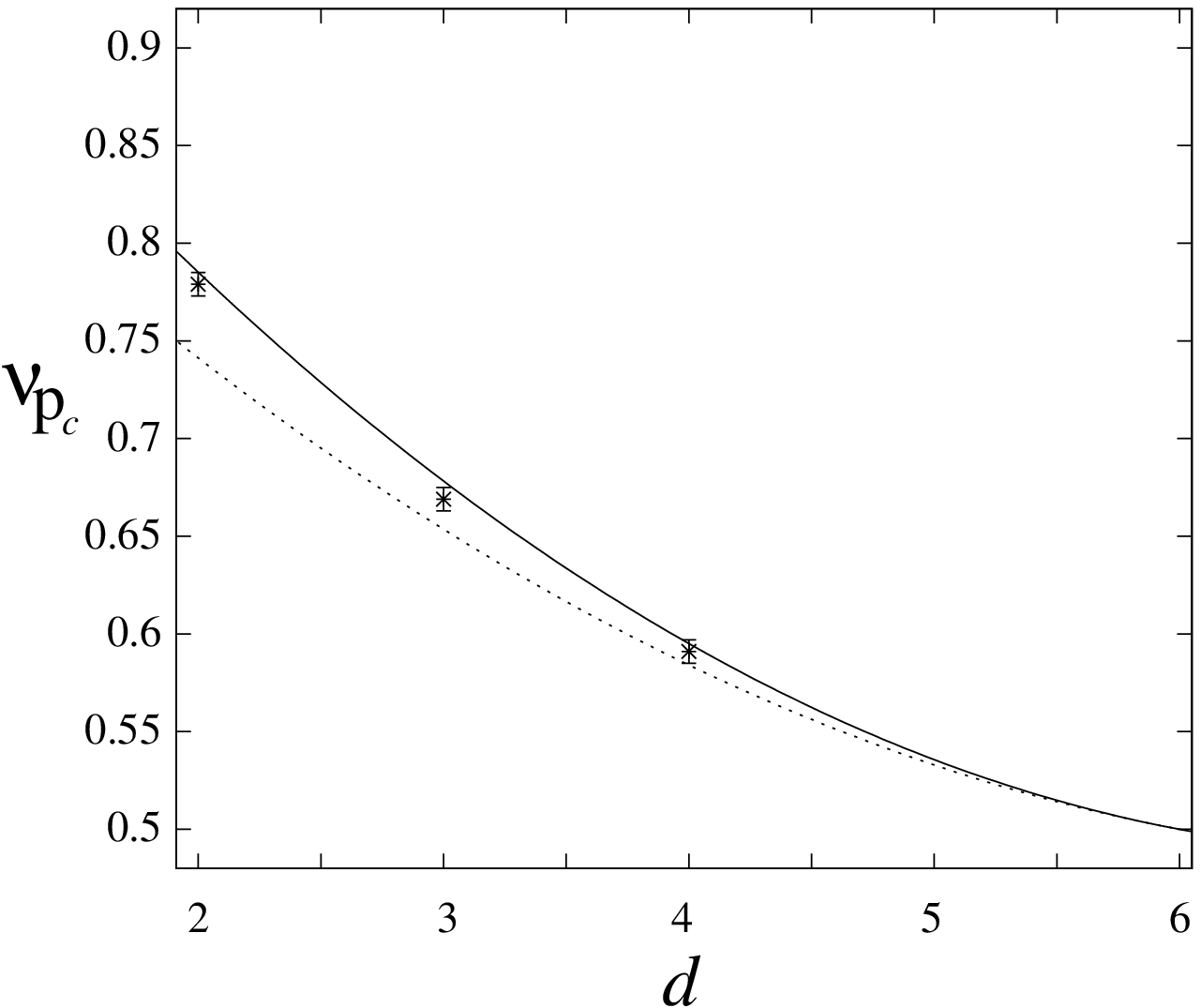}
\hspace*{0.5cm}
\includegraphics[width=5.9cm]{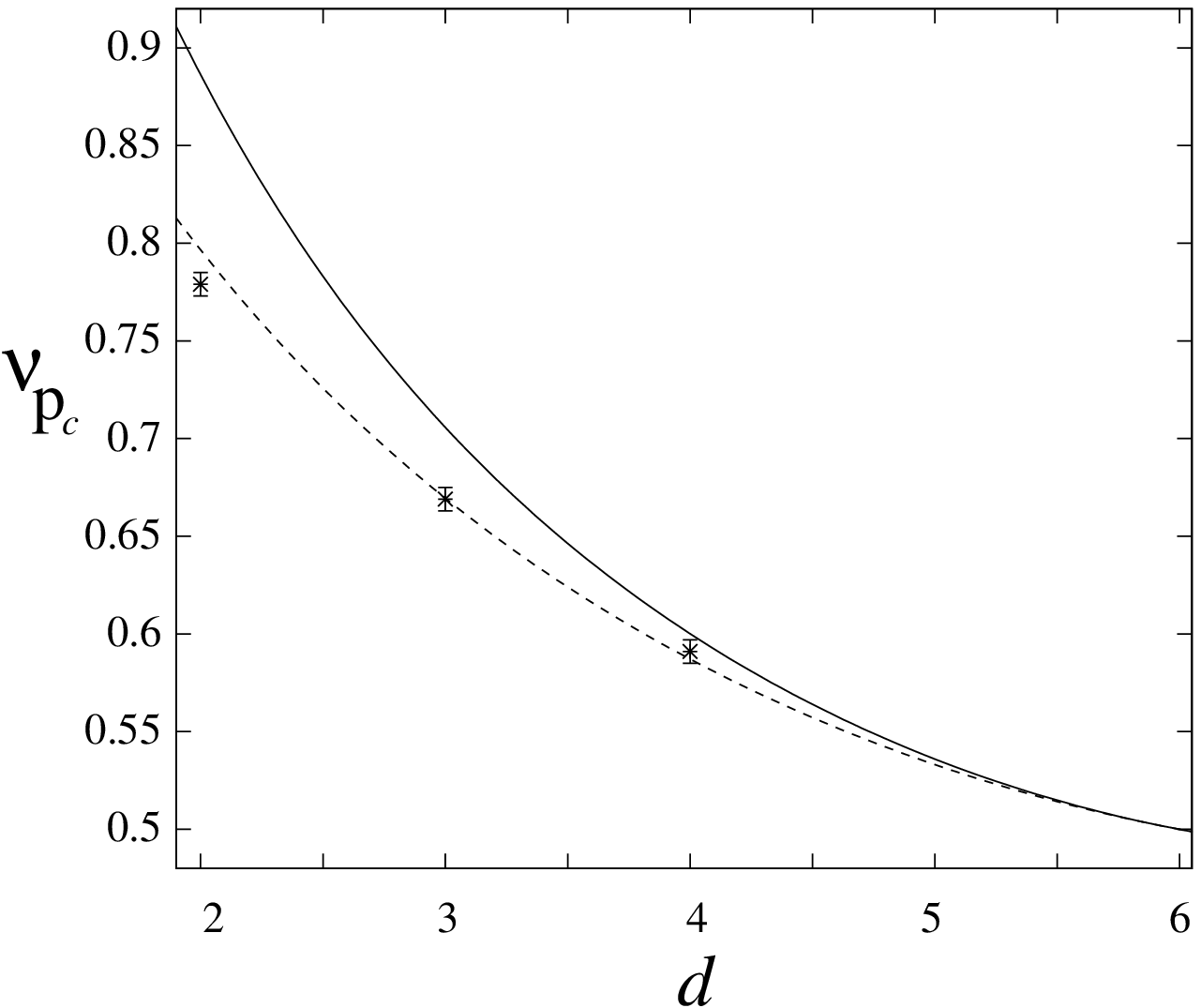}
\end{center}
\caption{\label{nusaw} Critical exponent $\nu_{p_c}$ of SAWs on percolation clusters as function of space dimension $d$; stars: numerical data, obtained by us. Left: Dotted line: analytical results 
of Janssen and Stenull (Eq.~(\ref{spectr})), solid line: analytical results 
of Ref.~\cite{Blavatska04}. Right: $[1]/[2]$ Pad\'e approximants of the same analytical results.}
 \end{figure}
 
The multifractal moments $M^{(q)}$ are defined as follows:
\begin{equation}
M^{(q)}=\sum\limits_{i \in R} {w(i)}^q.
\end{equation}
Averaged over different configurations of the constructed backbones of percolation clusters, they scale as:
\begin{equation} 
{\overline{M^{(q)}}}\sim R^{1/\nu^{(q)}},
\label{momexp}
\end{equation}
with exponents $\nu^{(q)}$  that do not depend on $q$ in  a linear or affine fashion, implying that SAWs on percolation clusters are multifractals.
To estimate the numerical values of  $\nu^{(q)}$ on the basis of data obtained by us, 
linear least-square fits are used.

\section{RESULTS}

At $q=0$ we just count the number of sites of the cluster of linear size $R$, and thus $1/\nu^{(0)}$ corresponds to the fractal 
dimension of the backbone $d_{p_c}^B$. Our results give $d_{p_c}^B(d{=}2)=1.647\pm0.006$, $d_{p_c}^B(d{=}3)=1.865\pm0.006$, $d_{p_c}^B(d{=}4)=1.946\pm0.006$.  
At $q=1$, we restore the value of the exponent $\nu_{p_c}$, governing the scaling law of the end-to-end distance for SAWs on the backbone of percolation 
clusters. We obtain $\nu^{(1)}(d{=}2)=0.779\pm0.006$, $\nu^{(1)}(d{=}3)=0.669\pm0.006$, $\nu^{(1)}(d{=}4)=0.591\pm0.006$, in perfect agreement with our recent numerical estimates 
\cite{Blavatska04,1,2,3}
based on the scaling of the end-to-end distance with the number of SAW steps. Comparison of our results for $\nu_{p_c}$ to that of the 
analytical studies \cite{Blavatska04,Janssen07} are presented in Fig. \ref{nusaw}.
 
\begin{figure}[t!]
 \begin{center}
\includegraphics[width=5cm]{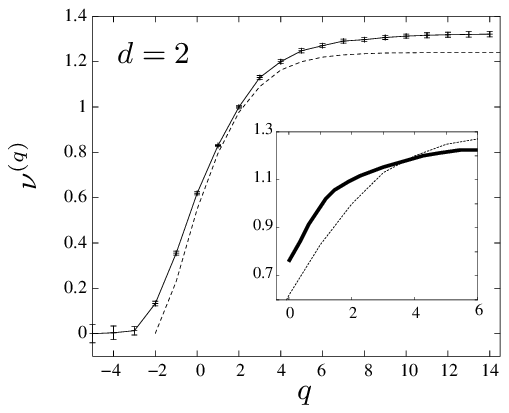}
\hspace*{0.02cm} 
\includegraphics[width=5.1cm]{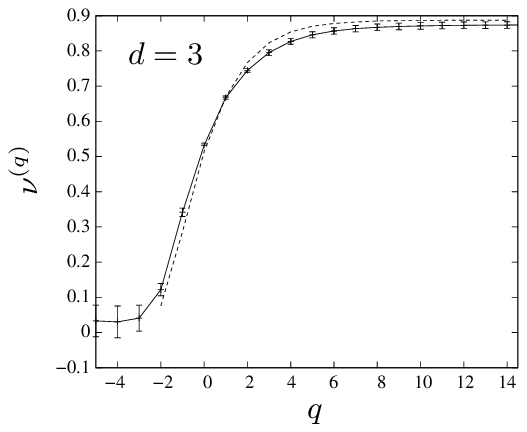}
\hspace*{0.02cm} 
\includegraphics[width=5.1cm]{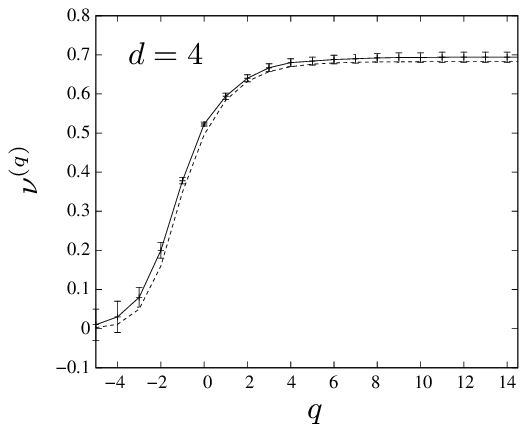}
\end{center}
\caption{\label{nuq} Spectrum of multifractal exponents $\nu^{(q)}$ as function of $q$ in $d=2, 3, 4$. The 
dotted lines present $[1]/[2]$ Pad\'e approximants to the analytical results 
of Janssen and Stenull (Eq.~(\ref{spectr})). The inset for $d=2$ shows a comparison with results 
from Ref.~\cite{Duplantier99} (bold solid line). }
 \end{figure}

The precision of our estimates decreases with increasing $q$. This problem turns out to be also especially 
crucial when exploring the moments with negative powers $q$: the sites with small probabilities to be visited, which are determinant in negative moments, are very difficult to probe.  

Our estimates of the exponents $\nu^{(q)}$ for different $q$ are presented in Fig.~\ref{nuq}. These values appear to be in perfect correspondence 
with analytical estimates down to $d=2$ dimensions, 
derived by  applying Pad\'e  approximation to the $\varepsilon=6-d$-expansion (\ref{spectr}), presenting the given series as 
ratio $[m]/[n]$ of two polynomials of degree $m$ and $n$ in $\varepsilon$.
We used the $[1]/[2]$ approximant, because it appears to be most reliable in restoring the known 
estimates in the limiting case $q=0$. A direct use of the  expression (\ref{spectr}) gives worse results, especially for low dimensions $d$ where the 
expansion parameter $\varepsilon=6-d$ is large.

\section{CONCLUSIONS}

To conclude, we have shown numerically that SAWs residing on the backbone of percolation clusters give rise to a whole spectrum of singularities, 
thus revealing multifractal 
properties. To completely describe peculiarities of the model, the multifractal scaling should be taken into account. We have found estimates for 
the exponents, 
governing different moments of the weight distribution, which scale independently.

\section*{ACKNOWLEDGMENTS}
Work supported by EU Marie Curie RTN ``ENRAGE'': {\em Random Geometry
and Random Matrices: From Quantum Gravity to Econophysics\/} under grant
No.\ MRTN-CT-2004-005616, S\"achsische DFG-Forschergruppe FOR877,
DFG Graduate School of Excellence ``BuildMoNa'', and Research Academy
Leipzig (RAL). V.B. is grateful for support through the ``Marie Curie 
International Incoming Fellowship" EU Programme and a Research Fellowship 
of the Alexander von Humboldt Foundation.

\end{document}